# DEPTH-SENSING INDENTATION ON $REBA_2CU_3O_{7-\delta}$ SINGLE CRYSTALS OBTAINED FROM XENOTIME MINERAL

Francisco Carlos Serbena*[1], Carlos Eugênio Foerster[1], Alcione Roberto Jurelo[1], Alexandre Mikowski[2], Pedro Rodrigues Júnior[1], Célia Regina Carubelli[3] and Carlos Maurício Lepienski[4]

[1] Departamento de Física, Universidade Estadual de Ponta Grossa, Av. Gen. Carlos Cavalcanti 4748, Campus de Uvaranas, CEP 84.030-000, Ponta Grossa, Paraná, Brazil

[2] Centro de Engenharia da Mobilidade, Universidade Federal de Santa Catarina, Campus Universitário, Bairro Bom Retiro, Caixa Postal 246, CEP 89.219-905, Joinville, Santa Catarina, Brazil

[3] Departamento de Química, Universidade Estadual de Ponta Grossa, Av. Gen. Carlos Cavalcanti 4748, Campus de Uvaranas,CEP 84.030-000, Ponta Grossa, Paraná, Brazil

[4] Departamento de Física, Universidade Federal do Paraná, Cx. Postal 19044, CEP 81531-990, Curitiba, Paraná, Brazil

* Corresponding author: fserbena@uepg.br, Phone: +55 42 3320 3044, Fax: +55 42 3220 3042

## Abstract

A natural mixture of heavy rare earths oxides extracted from xenotime mineral have been used to prepare large single crystals of high-temperature $REBa_2Cu_3O_{7-\delta}$ superconductor grown using the CuO-BaO self-flux method. Its mechanical properties along the *ab*-plane were characterized using instrumented indentation. Hardness and elastic modulus were obtained by the Oliver and Pharr method and corresponds to 7.4 ± 0.2 GPa and in range 135-175 GPa at small depths, respectively. Increasing the load promotes the nucleation of lateral cracks that causes a decrease in hardness and the measured elastic modulus by instrumented indentation at higher loads. The indentation fracture toughness was estimated by measuring the radial crack length from cube-corner indentations at various loads and was 0.8 ± 0.2 $MPa.m^{1/2}$. The observed slip systems of $REBa_2Cu_3O_{7-\delta}$ single crystals were [100](001) and [010](001), the same as for $YBa_2Cu_3O_{7-\delta}$ single crystals. The initial stages of deformation and fracture in the indentation process were investigated. The hardness and elastic modulus were not strongly modified by the crystallographic orientation in the *ab*-plane. This was interpreted in terms of the resolved shear stresses in the active slip systems. Evidence of cracking along the {100} and {110} planes on the *ab*-plane was observed. As a conclusion, the mechanical properties of $REBa_2Cu_3O_{7-\delta}$ single crystals prepared from xenotime are equivalent to those of $YBa_2Cu_3O_{7-\delta}$ single crystals produced by conventional rare earths oxides.

## 1. Introduction

Rare earth elements play an important role in energy production technologies such as high-energy density materials and high-temperature superconductors generators. Recently, they are becoming increasingly more expensive due to limited global supply with escalating costs for energy technologies that use them. Yttrium has been nominated among one of the most important rare earth elements for clean energy applications and with highest supply risk in the medium term [1]. Therefore, there is a demand for new technologies and materials that promote a reduction in cost production by developing alternatives to the conventional methods and with comparable performance.

It is known that minerals such as monazite, bastnäsite and xenotime are sources for rare earth elements such as the lanthanides and Yttrium [2]. Particularly, xenotime is a mixture of phosphates of Yttrium and heavy lanthanides (from Gadolinium to Lutetium) from which a natural mixture of rare earth trivalent oxides and Y can be extracted. This provide an alternative non-expensive route to prepare high-$T_c$ superconductor isostructural to $YBa_2Cu_3O_{7-\delta}$ that does not demand high purity Yttrium or other pure lanthanide element on Y site. Polycrystalline, melt-textured and single crystals samples using rare earth elements extracted from xenotime has been successfully grown with good superconductor properties [3-5].

The growth of large and good quality single crystals of high-temperature superconductors is necessary because it is important to measure their physical properties with high accuracy as well as to develop new devices. Various methods have been described for the growth of RE-123 (RE = Y or any lanthanides elements except Ce, Pr, Pm and Tb) single crystals. One of them is the flux method introduced by Schneemeyer et al. [6] and Wang et al. [7] using a CuO-BaO self-flux to grow $YBa_2Cu_3O_{7-\delta}$ single crystals. The main advantage of this method is that it allows the growth of lengthwise single crystals and that are easily extracted from the solidified flux matrix. In addition to its electrical and magnetic properties, the superconductors must also have good mechanical performance for technological applications.

The mechanical properties of high-temperature superconductors can be improved by several techniques. One of them is the addition of Ag, which causes the reduction of pores and cracks in the microstructure [8-9], higher tensile and bending strengths [10] and an increase in microplasticiy [11]. Another is in melt-textured samples, where the incorporation of particles of the $Y_2BaCuO_5$ phase and Ag doping in the matrix increases hardness, elastic modulus and fracture [8,12-13 ].

In a previous work [3], the hardness and elastic modulus of single crystals of $REBa_2Cu_3O_{7-\delta}$ superconductor successfully grown using rare earth oxides extracted from xenotime ore were similar than those measured for fully oxygenated $YBa_2Cu_3O_{7-\delta}$ single crystals and for melt-textured $YBa_2Cu_3O_{7-\delta}$. However, the only expected source of strengthening in $REBa_2Cu_3O_{7-\delta}$ single crystals is doping by the different ion species from the xenotime mineral and it would be expected the hardness of single crystals prepared from the xenotime mixture to be higher than the pure $YBa_2Cu_3O_{7-\delta}$ single crystals or melt-textured $YBa_2Cu_3O_{7-\delta}$. Another possibility was indentation fracture that could affected hardness in the *ab*-plane.

Therefore, we investigate in this work in much more detailed the mechanical properties (hardness, elastic modulus and indentation fracture toughness) of $REBa_2Cu_3O_{7-\delta}$ single crystals by instrumented indentation testing. Observation of the active slip systems and the mechanical properties variations with different crystallographic orientations as well as theoretical modeling of the hardness anisotropy were performed in order to understand the possible influences of solid solution hardening and indentation fracture on hardness.

## 2. Experimental Procedure

The chemical procedure to obtain the natural mixture of trivalent rare earth oxides $RE_2O_3$ from xenotime mineral is described in detail elsewhere [5]. A CuO-BaO self-flux method similar to that described by Schneemeyer *et al.* [6] and Wang *et al.* [7] was performed in order to grow large $REBa_2Cu_3O_{7-\delta}$ single crystals. The main advantage of this method is that it allows the growth of lengthwise single crystals and that are easily extracted from the solidified flux matrix.

The starting material was a powder mixture of $RE_2O_3$, $BaCO_3$ (99.9% purity) and CuO (99.99%) at a molar ratio of 1RE.4Ba.10Cu. This means that the starting mixture had a relative excess of Ba and Cu in comparison to the final desired composition of $REBa_2Cu_3O_{7-\delta}$. This mixture was ground during 4 hours in a ball mill and then placed in a $ZrO_2$ crucible that was heated up to 980°C at a 100°C/hour rate in air. After 1 hour at this temperature, a slow cooling process (3°C/h) began up to 880°C followed by furnace cooling to room temperature. At the end of this process, grown-up single crystals were extracted and oxygenated at 420°C during 5 days in flowing $O_2$ atmosphere in order to yield the superconductive orthorhombic structure. Large single crystals with typical size 3 x 3 x 0.03 $mm^3$ were obtained. These samples were characterized by XRD, SEM and electrical resistivity and the results were described elsewhere [5].

Hardness (*H*) and elastic modulus (*E*) profiles were determined using a Nanoindenter XP™ with a diamond Berkovich indenter and following the Oliver & Pharr method [14]. The indentations were performed in 10 loading-unloading cycles up to a maximum load of 400 mN. Several indentations were performed at different samples and at different regions of each sample to check for sample homogeneity. Because the surface available for indentations in each sample was small, the indentations were performed in grids of 3 x 3 or 2 x 2 at several places of each sample. The distance between the indentations was at least 50 μm.

The effect of load on the fracture pattern produced by the indentation was studied by performing numerous indentations at different loads using a cube-corner indenter on the *ab*-plane. The crack pattern was observed by SEM images using backscatter or secondary electrons.

Room temperature indentation fracture toughness $K_C$ was obtained using a diamond cube-corner indenter at single loading/unloading cycles. The maximum load was 400 mN. $K_C$ was calculated according to the relation proposed by Harding *et al.* [15] for a cube-corner indenter:

$$K_C = \alpha.\sqrt{\frac{E}{H}}\frac{P}{c^{3/2}} \quad (1)$$

where $\alpha$ is a constant equals to 0.036 ± 0.014, *c* is the radial crack length and *P* is the applied load. The crack lengths were measured by optical microscopy and SEM as soon as possible after the indentations tests in order to avoid chemio-mechanical deleterious effects [16-18.]

The initial stages of deformation and fracture in the indentation process were also investigated by performing single loading-unloading cycles using a cube-corner indenter at a load of 20 mN. This load was used in order to avoid extensive cracking that occurs at higher loads. The cube-corner indenter was chosen because it promotes cracking at smaller loads than the Berkovich indenter and the strains induced are much higher. The influence of crystallographic orientation was investigated by rotating the sample from 0° to 60° at 15° steps between tests. At least 15 indentations were performed for each orientation. Hardness using the cube-corner indenter was calculated by the ratio *P/A*, where *P* is the maximum load and *A* is the projected area of the hardness impression. *A* is calculated from the contact depth $h_c$ as a polynomial function $C_0h_c^2 + C_1h_c + C_2h_c^{1/2} + \ldots$ where $C_i$ are constants. For a perfect cube-corner indenter, $A = 2.598.h_c^2$. The elastic modulus was also calculated from the initial unloading curve using the Oliver-Pharr method and using 50% of the unloading curve. Both the Berkovich and cube-corner indenters' area functions and frame compliance were calibrated using a standard fused silica sample.

The [100] and [010] directions on the *ab*-plane were determined from the observation of the orientation of the growth steps on the sample surface.

## 3. Results

Figs. 1 shows typical loading-unloading curves for the *ab*-plane using a Berkovich tip indenter. Contact depths between 2200 nm and 2800 nm are obtained at a load of 400 mN. Several discontinuities are observed along the curve, especially at contact depths greater than around 600 nm. These discontinuities are associated with crack nucleation which are parallel to the surface and are nucleated at different loads as can be observed in figs 1(b) and 1(c), sometimes with a large crack as highlighted in fig. 1(b). Due to the lamellar structure of the $REBa_2Cu_3O_{7-\delta}$ structure, these cracks are lateral that once nucleated promote detachment of the material under the indenter (chipping). Then, the indenter loses support and a discontinuity takes place in the loading curve.

Slip lines parallel to the [100] and [010] directions are observed inside the indentations (fig. 2). These are the traces of the [100](001) and [010](001) slip systems reported in the literature for $YBa_2Cu_3O_{7-\delta}$ single crystals [19-20], indicating the slip systems for $REBa_2Cu_3O_{7-\delta}$ and for $YBa_2Cu_3O_{7-\delta}$ are the same. It is important to note that these slip systems do not produce slip lines on the *ab*-plane, as they are parallel to it, but do produce inside the indentations.

Figs. 3(a) and 3(b) show the hardness and elastic modulus profiles as a function of the contact depth, respectively. The hardness is 7.4 GPa near the surface and remains constant up to a contact depth of 600 nm approximately. At larger contact depths, the hardness decreases, reaching a minimum of 4.6 GPa at maximum contact depth. The elastic modulus decreases continuously from 177 GPa near the surface to 115 GPa at a depth of 600 nm and reaches 60 GPa at the maximum contact depth.

Although the hardness may decrease with increasing indentation depth due to the indentation size effect [21], no decrease in elastic modulus should be observed. In order to obtain more information, we measured the area of each Berkovich indentation from the SEM photographs and calculated the hardness at maximum depth dividing the load of 400 mN by the measured area. The hardness measured in this manner was compared with those obtained by instrumented indentation and they agree. Consequently, the low hardness measured at maximum depth is characteristic of $ReBa_2Cu_3O_{7-\delta}$ single crystals due to nucleation and propagation of lateral cracks at loads above a certain threshold. It is expected these cracks also affected the measured contact stiffness and the calculated elastic modulus as well.

The cracking dependence with load for cube-corner indenter is shown in fig. 4. The cracking pattern for 20mN, 200mN and 400 mN can be observed. At 20 mN, well delineated radial cracks aligned along the indenter diagonals are observed in fig. 4(a). There is also material displaced at the edges of the imprint. The displacement appears to be lamellar, following the lamellar crystalline structure of the orthorhombic unit cell. As the load increases, the damage around the indentation intensifies. Lateral cracks are observed and their extent increases with load. At 200 mN (fig. 4(b)), they affect the observation of radial cracks and it is not possible to the use their length to calculate the indentation fracture toughness. At the maximum load of 400 mN (fig. 4(c)), the damage is so severe that lateral cracks propagate completely around the indentation impression and parallel to the surface, extending beyond the radial cracks and promoting an chipping.

As the radial cracks were not well defined at high loads, only those produced up to a load of 100 mN were used to estimate the indentation fracture toughness of the *ab*-plane of single crystals. Table 1 shows the indentation fracture toughness $K_C$ calculated using eq. (1) as well as the average crack length *c* measured for each load. The indentation fracture toughness is constant, does not depending on load, and a value of 08 ± 0.2 MPa.m$^{1/2}$ is obtained.

The anisotropy in the mechanical properties is shown in figs. 5(a) and 5(b). Hardness and elastic modulus measured using a cube-corner indenter are plotted as a function of the orientation $\theta$ of the cube-corner indenter diagonal from the assumed [100] direction in the *ab*-plane. There is no variation with crystallographic orientation within experimental error both for *H* and *E*. The measured hardness is around 9 GPa and the elastic modulus is 160 GPa using the cube-corner indenter. This hardness is slightly higher and the elastic modulus is approximately the same than the hardness and the elastic modulus measured using the Berkovich indenter. Differences in hardness have been observed in some materials when the hardness is obtained using a Berkovich or a cube-corner indenter [22]. It has also been observed that the use of both indenters results in the same elastic modulus.

Fig. 6 shows the plastic deformation around the indentations when the direction of the cube-corner diagonals is varied on the *ab*-plane. The anisotropy in plastic deformation around the indentation is evident, although no anisotropy in hardness and elastic modulus was measured. We observed two slip directions perpendicular to each other at the surface. These directions are the <100> directions, which corresponds to the activation of the [100](001) and [010](001) slip systems observed in $YBa_2Cu_3O_{7-\delta}$ single crystals and are marked by arrows in Figs. 6(a), 6(b), 6(d) and 6(e). No slip lines are observed since these slip systems are parallel to the *ab*-plane. It is also observed radial cracks partially aligned with the indenter diagonals along the <110> and <100> directions. It is reported in the literature the primary crack plane is along the {100} planes for $YBa_2Cu_3O_{7-\delta}$ single crystals [23-24]. In some orientations, lateral detachment of material around the indentation is observed, possibly as a result of lateral crack nucleation as observed in figs. 6(c) and 6(e).

**4. Discussion**

The hardness using a Berkovich indenter was 7.4 ± 0.2 GPa and approximately constant near the surface. For a contact depth greater than 600 nm, it decreases continuously to 4.6 GPa at the maximum depth. The elastic modulus also decreased continuously from 177 GPa near the surface to 60 GPa at maximum depth. Several discontinuities in the loading curve were observed for indentation depths greater than 600 nm and they are associated with the nucleation of lateral cracks. This affects the measurement of hardness and elastic modulus in two ways: (a) as the crack is nucleated and material is displaced, the indenter penetrates further into material and a greater penetration depth is obtained; (b) the contact stiffness in the beginning of the unloading cycle of the indentation changes and causes a decrease in the calculated elastic modulus. These facts are further confirmed as the hardness measured by instrumented indentation and that obtained by direct measurement of the area were the same.

Table 2 compares the hardness and elastic modulus of various superconductor systems with different microstructures and compositions. Usually, the hardness of polycrystalline samples are lower than melt-textured and single-crystals samples. This is attributed to the existence of pores. The hardness of melt-textured samples is higher due to the lower porosity and the presence of the RE-211 phase. It is also observed the hardness of the *a(b)c*-plane is slightly lower than the *ab*-plane. The value measured in this work is in agreement with hardness measured by instrumented indentation on *ab*- or *a(b)c*-planes of melt-textured Y-123 samples which is in the range 7-11 GPa [25-26] and slightly lower than that measured in an earlier work of 8.5 ± 0.5 GPa on the *ab*-plane of $REBa_2Cu_3O_7$ single crystals [5]. The hardness also agrees with that measured on Ru-1212 polycrystalline samples, which has a structure very similar to $YBa_2Cu_3O_{7-\delta}$ and is around 8.6 GPa [27]. $REBa_2Cu_3O_{7-\delta}$ single crystals have small amounts of Dy, Ho and Er dissolved in its structure [5], probably in substitutional solid solution to Y. The ionic radius of Dy, Ho and Er are 0.912 A, 0.901 A and 0.881 A, respectively. The ionic radius of Y is 0.9 A. Consequently, mainly Dy and Er ions distort the lattice due to the different ionic radius, generate stresses and increase the barrier for dislocation motion. It was expected the presence of these ions would result in an increase in the hardness of $REBa_2Cu_3O_{7-\delta}$ single crystals by solid solution strengthening.

However, the fact that the hardness are the same indicates that another mechanism maybe operating. It was observed in Figs. 1 extensive lateral microcracks produced by the indentations due to the lamellar structure of the $REBa_2Cu_3O_{7-\delta}$ structure. They probably are nucleated at very load loads and displaces material underneath the indenter. This produces a larger tip penetration depth and a smaller contact stiffness. While a larger tip penetration decreases hardness proportional to $1/h_c^2$, where $h_c$ is the contact depth, a smaller contact stiffness increases hardness proportional to $h_c$ [14]. The overall result is a decrease in hardness with increasing penetration depth and this might explain the same measured hardness of $REBa_2Cu_3O_{7-\delta}$ and $YBa_2Cu_3O_{7-\delta}$ single crystals.

The continuous decrease of the elastic modulus with contact depth as well as its low value at high depths when compared with literature data is attributed to the crack nucleation produced by indentations. The elastic modulus measured at low loads up to a depth of 200 nm is in the range 135–175 GPa. This range is in agreement with those measured in other studies: the elastic modulus was in the range 160-175 GPa for melt-textured $Y_{0.95}Er_{0.05}Ba_2Cu_3O_{7-\delta}$ [25]; for melt-textured Y-123 samples doped with Ag, the elastic modulus was in the range 150-220 GPa. Meanwhile, for pure Y-123 melt-textured samples, the elastic modulus was between 160-170 GPa [26] and for pure polycrystalline samples was around 125 GPa [28]. This range agrees with that measured in an earlier work of 143 ± 3 GPa on the *ab*-plane of $REBa_2Cu_3O_7$ single crystals [5]. The elastic modulus of Ru-1212 polycrystalline samples was around 145 GPa [27].

The indentation fracture toughness was 0.8 ± 0.2 $MPa.m^{1/2}$ for our samples. This value is consistent with those reported in the literature [13,25-26,29] as measured by the indentation technique for high-temperature superconductors samples with different microstructures. The indentation fracture toughness of the *ab*-plane of melt-textured $REBa_2Cu_3O_{7-\delta}$ samples was in the range 0.8-1.2 $MPa.m^{1/2}$ [26]. $K_C$ was also measured in both *ab*- and *a(b)c*-planes of melt-textured $Y_{0.95}Er_{0.05}Ba_2Cu_3O_{7-\delta}$ samples and were 1.3 ± 0.3 and 0.8 ± 0.2 $MPa.m^{1/2}$, respectively [25]. Indentation fracture toughness was also measured in Ag-doped melt-textured Y-123 [26]. The values were higher, in the range 1.4-1.6 $MPa.m^{1/2}$ for both *ab*- and *a(b)c*-planes owing to the ability of Ag for filling pores and cracks in the microstructure. Fracture toughness measured by other methods such as using SENB (single edge notch bend) samples are around 1 $MPa.m^{1/2}$ for undoped samples [13]. The fracture toughness for the *ab*- and *a(b)c*-planes are 0.8 and 0.32 $MPa.m^{1/2}$, respectively [29].

Foerster et al. [30] compared the mechanical properties of samples with different microstructures produced by using the elements extracted from xenotime. It was observed that the highest measured hardness and elastic modulus are of the melt-textured samples because of their microstructure of 211-inclusions and pores embedded in the 123-matrix. The lowest hardness is observed for the polycrystalline sample due to its high porosity. Indentation fracture toughness for melt-textured samples is in the range of literature values (0.7-1.0 $MPa.m^{1/2}$). It was concluded that melt-textured $REBa_2Cu_3O_{7-\delta}$ samples have the most favorable mechanical properties for technological applications.

We observed no variation of *H* and *E* with crystallographic orientation in the *ab*-plane within experimental error and this indicates that the mechanical properties are not strongly modified by the crystallographic orientation. However, anisotropy in plastic deformation was observed. For a better understanding, we used the model of Brookes et al. [31] to predict what are the stresses on them and how the hardness varied with the indenter orientation. According to this model, the hardness anisotropy is determined by the active slip systems that accommodate plastic deformation produced in the hardness test by dislocation motion. Therefore, materials with the same active slip systems show similar anisotropic hardness.

The model assumes a few assumptions: a) the maximum tensile stress acting on the displaced material is along an axis parallel to the steepest slope along the face of the indenter represented by the F'F' direction in Fig. 7(a); b) the possibility of rotation of the slip system favors its activation; c) the slip system that is able to rotate about an axis parallel to indenter face and the surface is most favorable activated (when directions a.r. and HH in Fig. 7(b) are coincident and $\psi$ = 0) and d) the maximum constraint on rotation of the slip system is when the slip direction and the direction HH are coincident ($\gamma$ = 0 in Fig. 7(b)).

According to this model, the effective resolved shear stress $\tau_e$ on a particular slip system is:

$$\tau_e = \frac{1}{2}\left(\frac{F}{A}\right)\cos\lambda.\cos\phi.\left(\cos\psi + \sin\gamma\right) \qquad (2)$$

where $F$ is the applied load, $A$ is a cross-sectional area, $\lambda$ is the angle between the slip direction and the stress axis considered as the steepest slope of a indenter face, $\phi$ is the angle between the normal to the slip plane and the stress axis, $\psi$ is the angle between the axis of rotation of the slip system and a direction parallel to the indenter face and the surface defined by HH and $\gamma$ is the angle between the slip direction and HH as shown in Figs. 7(a) and 7(b). The term $\cos\lambda.\cos\phi$ is related to the projection of the maximum tensile stress along FF on the slip plane and slip direction and the term ($\cos\psi$ + sin $\gamma$) is related to slip system rotation in relation to the HH direction as described in the preceding paragraph.

The ratio *F/A* is related to the maximum stress acting parallel to the indenter face and is not the material hardness. Therefore, we define a normalized effective resolved shear stress (ERSS) as the ratio $\tau_e/(F/A)$ and its computation allows the determination of the relative magnitudes of the shear stresses on different slip systems as a function of indenter diagonal on a particular crystal plane. Particular directions with higher shear stresses are the directions with lower hardness.

The observed slip systems are the same as those for Y-123 samples: (001) planes with [100] and [010] Burgers vectors [19-20]. For a given orientation $\theta$ of the indenter on the *ab*-plane, we calculated the ERSS using eq. (2) for a particular indenter face for both slip systems using vector calculus. The angle between the one of the faces of the cube-corner indenter and the central axis is 35.3$^{o}$ as shown in Fig. 7(a).

Fig. 8(a) shows the normalized effective resolved shear stress calculated for the two slip systems on a single face of a cube-corner indenter (grey face at the inset) as it rotates from the [100] direction on the *ab*-plane. Both slip systems operate simultaneously. From the geometry of both slip systems, the axis of rotation of both slip systems is always perpendicular to the *ab*-plane, then the angle $\psi$ is always 90$^{o}$ and the term $\cos\psi$ is zero in eq. (2). Therefore, ERSS is solely determined by the resolved shear stresses on the slip system and its ability to rotate due to the constraint term sin$\gamma$. When the direction HH of a particular face of the indenter is coincident with one of the slip directions [100] or [010], ERSS is zero. Shear stresses are zero at $\theta$ = 30°, 120°, 210° and 300°. This can be observed in fig. 6(c) as the material is less displaced at the face with zero stress.

Fig. 8(b) shows the average normalized effective resolved shear stresses considering the maximum value of shear stress acting on each of the three faces of the cube-corner indenter. Also plotted are the absolute stresses on the slip systems considering a single face for comparison. There is no significant variation of the average shear stress with orientation. This result can be understood as when in one face the shear stress is low (or even zero), the stresses are high in the other faces, resulting in a low variation of the average stress. Simulations were also performed for a Berkovich indenter and the same resolved shear stresses curves are observed but with a higher value. The calculations confirm the inexistence of hardness anisotropy on the *ab*-plane. The observed hardness anisotropy is consistent with the activation of (001)[100] and (001)[010] slip systems for $REBa_2Cu_3O_{7-\delta}$ single crystals, which are the same for $YBa_2Cu_3O_{7-\delta}$ single crystals.

## 5. Conclusions

The mechanical properties of the *ab*-plane of high-temperature $REBa_2Cu_3O_{7-\delta}$ superconductor were characterized by instrumented indentation. The hardness and elastic modulus were 7.4 ± 0.2 GPa and in the range 135-175 GPa at small depths, respectively. Increasing the load promotes crack nucleation that decreases the measured hardness and elastic modulus for higher depths. Indentation fracture toughness was measured using the radial crack length from cube-corner indentations and a value of 0.8 ± 0.2 MPa.$m^{1/2}$ was evaluated. The hardness and elastic modulus were not strongly modified by the crystallographic orientation in *ab*-plane. The activation of the (001)[100] and (001)[010] slip systems was predicted from observation of slip lines in the *ab*-plane and from hardness anisotropy theoretical modeling. The slip systems of $REBa_2Cu_3O_{7-\delta}$ single crystals are the same as for $YBa_2Cu_3O_{7-\delta}$ single crystals. Evidence of cracking along the {100} and {110} planes on the *ab*-plane was noticed. It is concluded that the similar hardnesses resemble the same operative active slip systems for both $REBa_2Cu_3O_{7-\delta}$ and $YBa_2Cu_3O_{7-\delta}$ compounds and the effect of lateral cracking during indentation, which counterbalances the expected solid solution strengthening as a result of the different ion species from the xenotime mineral in $REBa_2Cu_3O_{7-\delta}$ single crystals.

## 6. Acknowledgements

This work was partially financed by the CNPq Brazilian Agency under contract $n^0$ 475347/01-3. We acknowledge Dr. A. L. Chinelatto for the SEM measurements.

## Table Captions

**Table 1.** Cube-corner indentation load, crack length and indentation fracture toughness as calculated using equation (1). The values of *H* and *E* used were 7.5 GPa and 142 GPa, respectively.

**Table 2.** Hardness and elastic modulus from the literature for different high-temperature superconductor samples.

**Figure Captions**

**Fig. 1.** (a) Typical loading-unloading curves of a Berkovich indentation in the *ab*-plane of a $REBa_2Cu_3O_{7-\delta}$ single crystal showing tip incursions during loading associated with crack nucleation. Indentation cracking associated with loading curves represented by (b) open and (c) full symbols are displayed.

**Fig. 2.** SEM of a 400 mN Berkovich indentation on the *ab*-plane. Slip lines inside the indentation corresponds to the traces of the [100](001) and [010](001) slip systems.

**Fig. 3.** (a) Hardness and (b) elastic modulus as a function of contact depth for *ab*-plane of $REBa_2Cu_3O_{7-\delta}$ single crystal using a Berkovich indenter.

**Fig. 4.** SEM micrographs of cube-corner indentations on *ab*-plane of $REBa_2Cu_3O_{7-\delta}$ single crystal with loads of (a) 20 mN, (b) 200 mN and (c) 400 mN

**Fig. 5.** (a) Hardness and (b) elastic modulus measured using a cube-corner indenter as a function of the diagonal orientation of the cube-corner indenter for the *ab*-plane of $REBa_2Cu_3O_{7-\delta}$ single crystal.

**Fig. 6.** SEM photograph of 20 mN cube-corner indentations on the *ab*-plane and with one of the indenter diagonals aligned at (a) $0^o$, (b) $15^o$, (c) $30^o$ (d) $45^o$ and (e) $60^o$ from the <100> directions indicated by arrows.

**Fig. 7.** (a) Schematic representation of a slip system and its orientation in respect of the face of a cube-corner indenter and (b) the definition of the angles used in eq. (2) to calculate the resolved shear stress acting on the slip system. HH is the direction of the edge of the particular indenter face, F′F′ is the direction of the steepest slope of the indenter face, FF is the stress axis, N is the direction normal to the slip plane and a.r. is the rotation axis of the slip system.

**Fig. 8.** (a) Normalized effective resolved shear stress calculated using eq. (2) for the slip systems of $YBa_2Cu_3O_{7-\delta}$ on a single face of a cube-corner indenter (the indenter grey face shown in the inset) and (b) average normalized effective resolved shear stress of the three faces of a cube-corner indenter on the *ab*-plane as a function of the diagonal orientation.

| $P$ (mN) | $c$ (μm) | $K_C$ (MPa.m$^{1/2}$) |
| --- | --- | --- |
| 20 | 2.8 ± 0.3 | 0.7 ± 0.1 |
| 50 | 4.7 ± 0.7 | 0.8 ± 0.2 |
| 100 | 8 ± 1 | 0.8 ± 0.2 |

Table 1

| Sample | *H* (GPa) | *E* (GPa) |
| --- | --- | --- |
| polycrystalline Y-123[13] | 3.4 ± 0.7 | 125 ± 10 |
| polycrystalline Y-123 doped with Ag [13] | 3.2 ± 0.6 | 140 ± 10 |
| polycrystalline RE-123[7] | 5.0 ± 2 | 133 ± 15 |
| polycrystalline $RuSr_2GdCu_2O_{8+d}$[12] | 9 ± 2 | 150 ± 30 |
| *ab*-plane of melt-textured Y-123[11] | 8 ± 1 | 170 ± 10 |
| *a(b)c*-plane of melt-textured Y-123[11] | 6.9 ± 0.3 | 160 |
| *ab*-plane of melt-textured Y-123 doped with Ag[11] | 8 ± 1 | 140-220 |
| *a(b)c*-plane of melt-textured Y-123 doped with Ag[11] | 7.1 ± 0.8 | 140-180 |
| *ab*-plane of melt-textured RE-123[7] | 9.0 ± 0.5 | 206 ± 10 |
| *a(b)c*-plane of melt-textured RE-123[7] | 9.0 ± 0.5 | 190 ± 10 |
| *ab*-plane of melt-textured $Y_{0.95}Er_{0.05}Ba_2Cu_3O_{7-\delta}$ [10] | 7.4 ± 0.6 | 158 ± 4 |
| *a(b)c*-plane of melt-textured $Y_{0.95}Er_{0.05}Ba_2Cu_3O_{7-\delta}$[10] | 7.2 ± 0.3 | 176 ± 6 |
| *ab*-plane of RE-123 single crystals[6] | 8.5 ± 0.5 | 143 ± 3 |
| *ab*-plane of $YBa_2Cu_3O_7$ single crystals[15] | - | 182 |
| *a(b)c*-plane of $YBa_2Cu_3O_7$ single crystals [15] | - | 143 |

Table 1

fig1

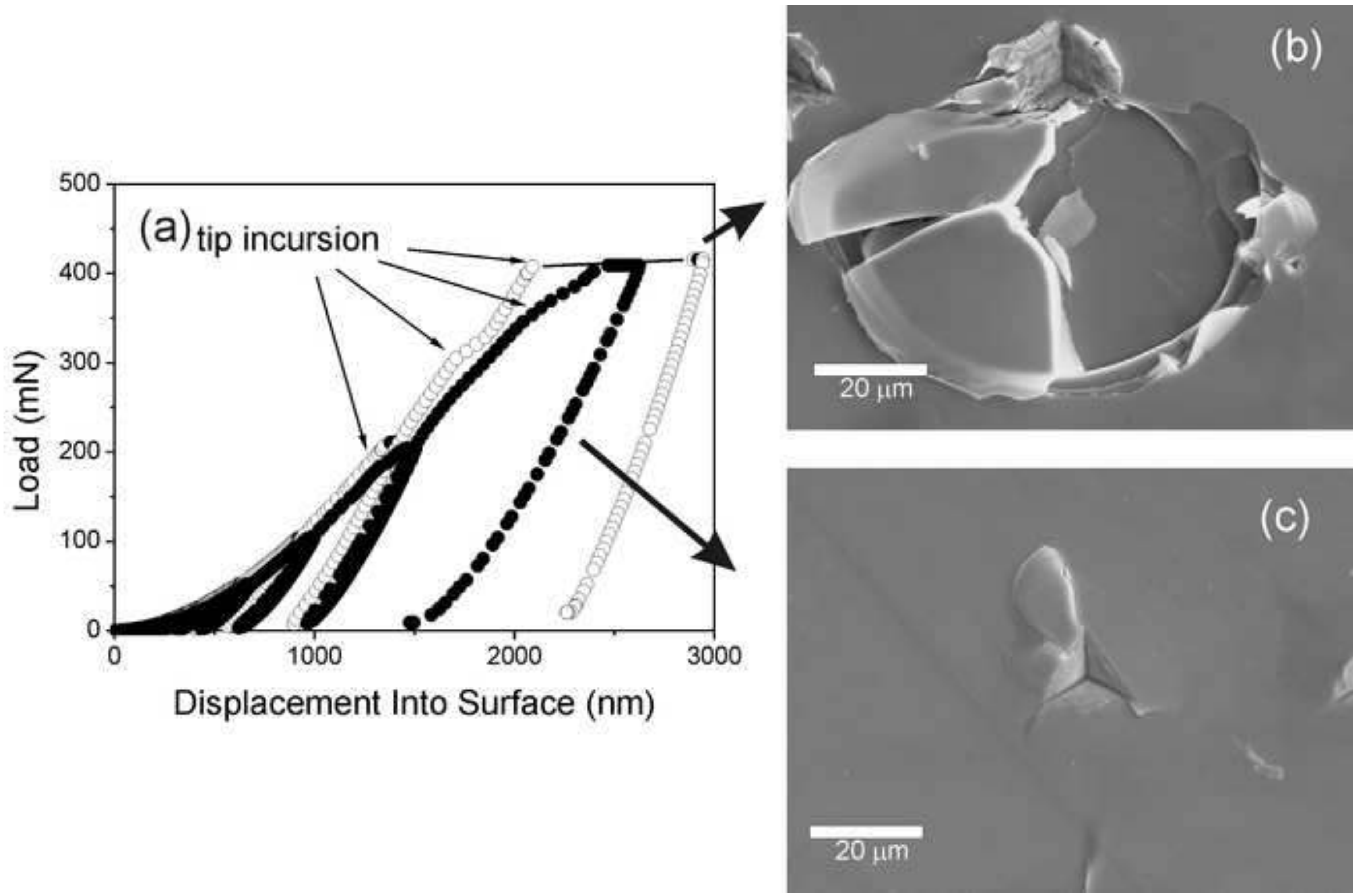

fig2
Click here to download high resolution image

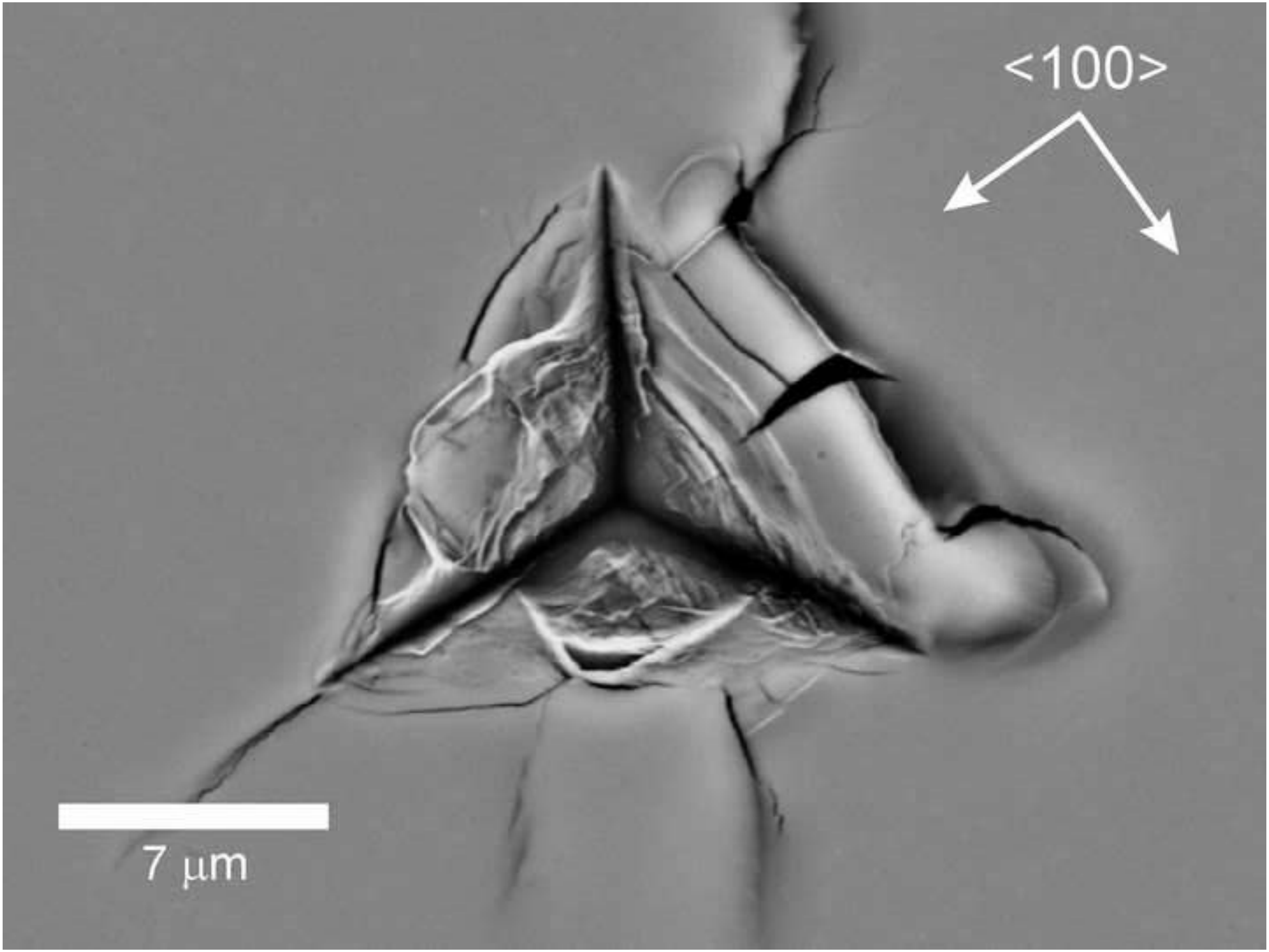

fig3
Click here to download high resolution image

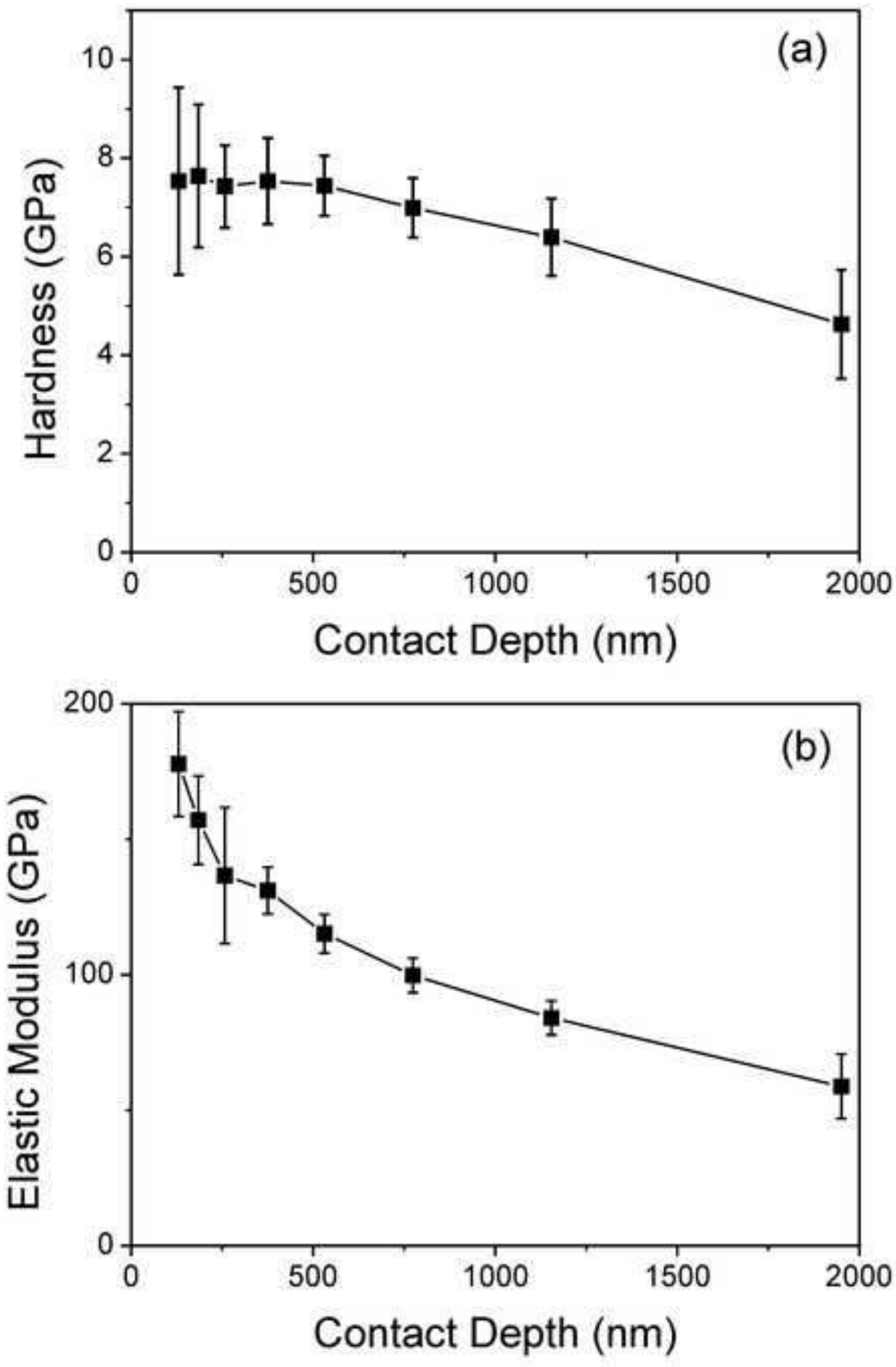

fig4
Click here to download high resolution image

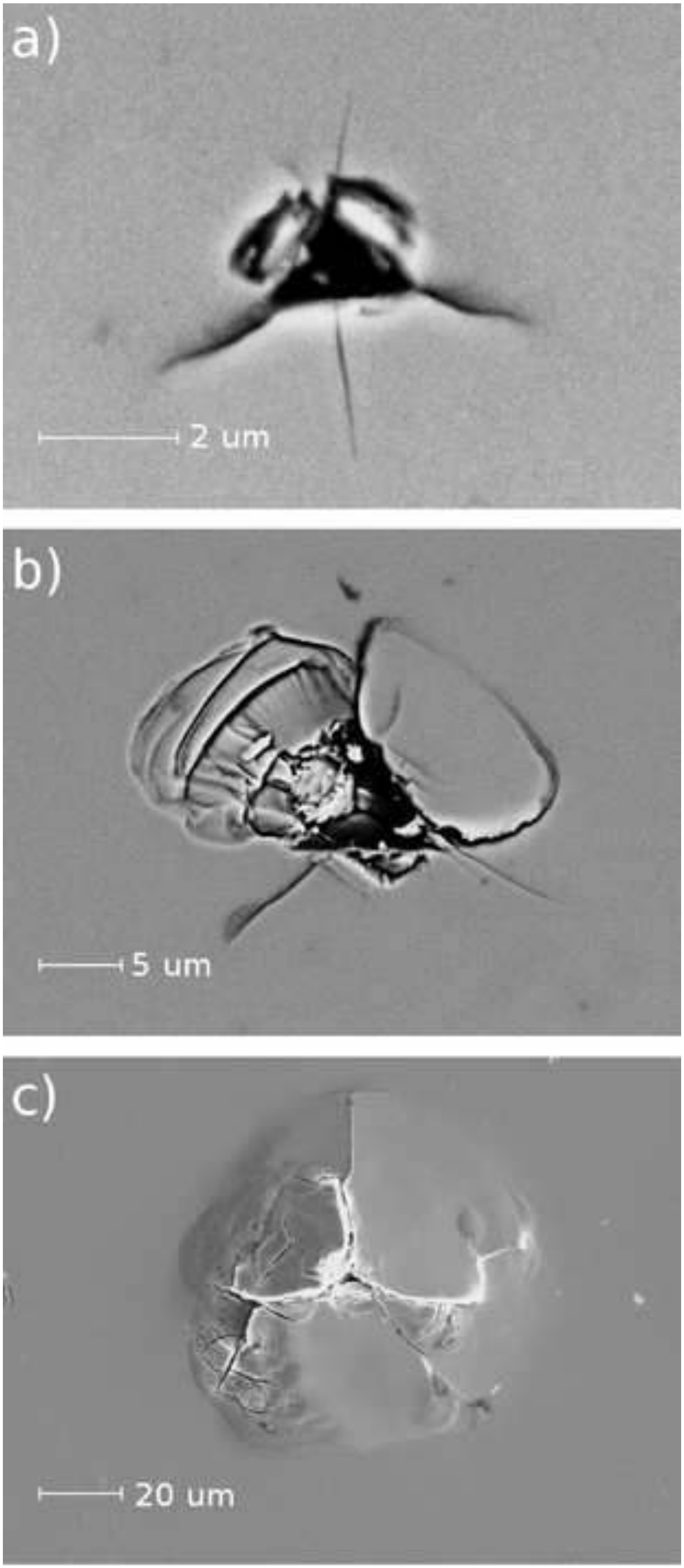

fig6
Click here to download high resolution image

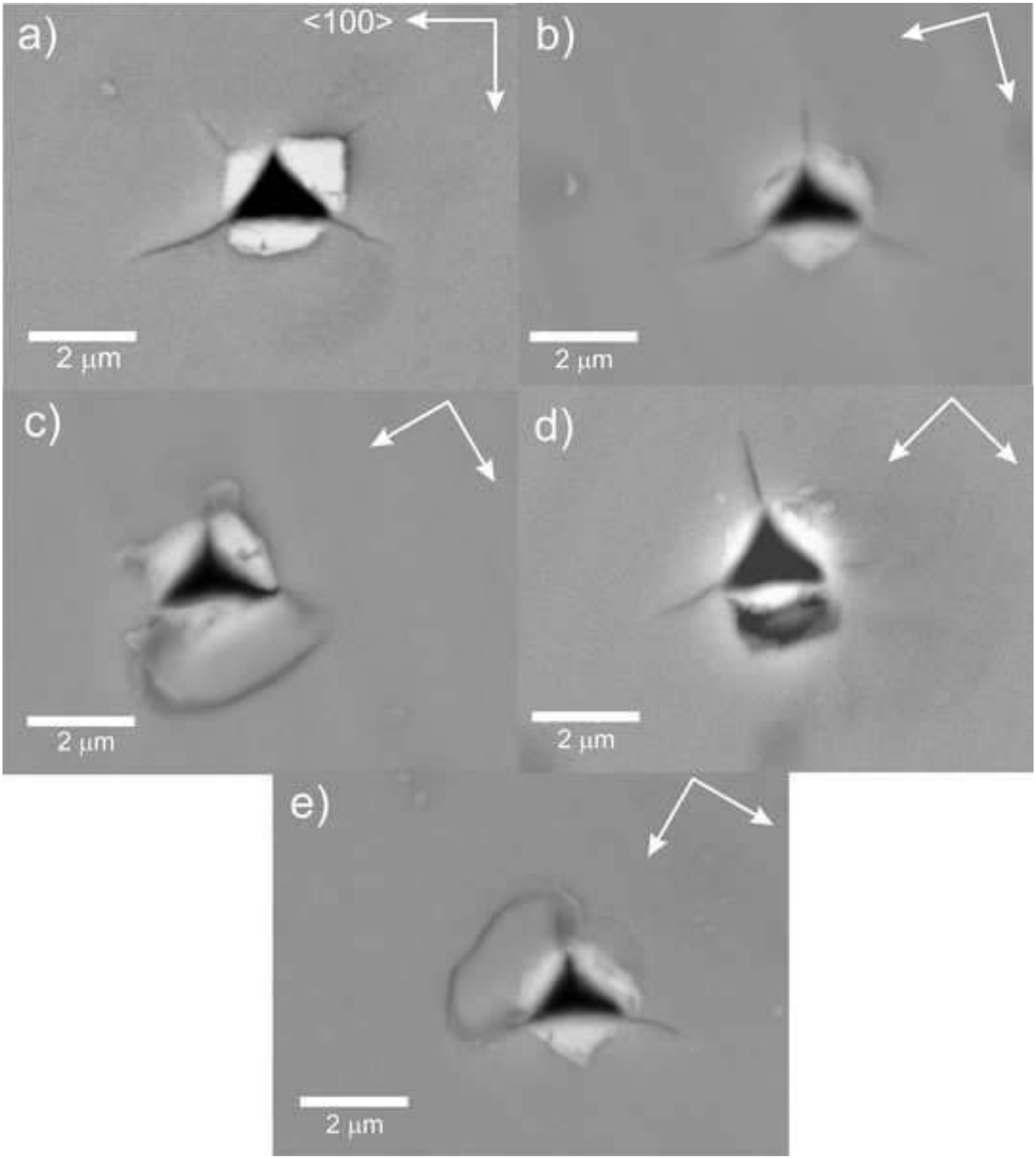

fig5
Click here to download high resolution image

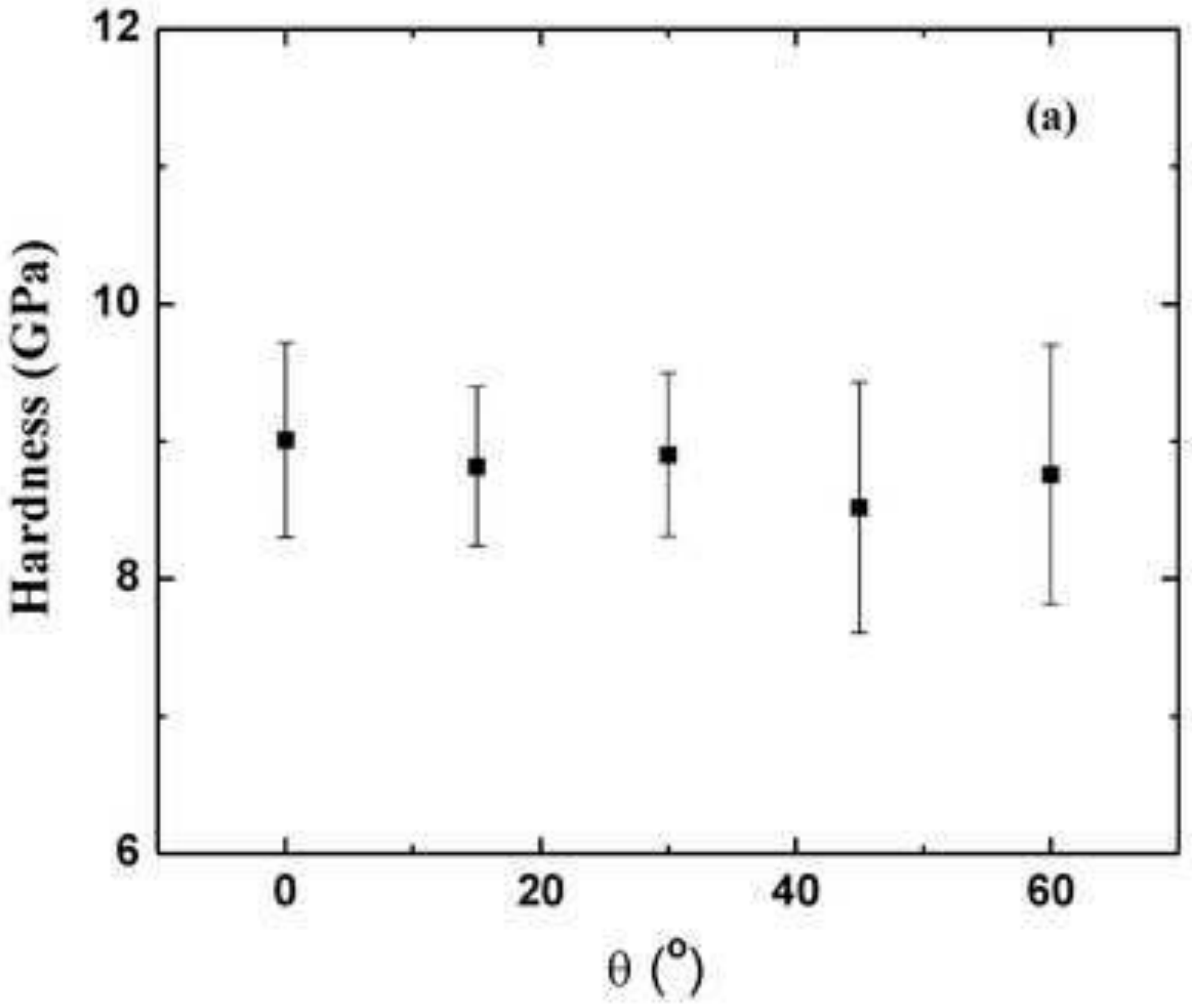

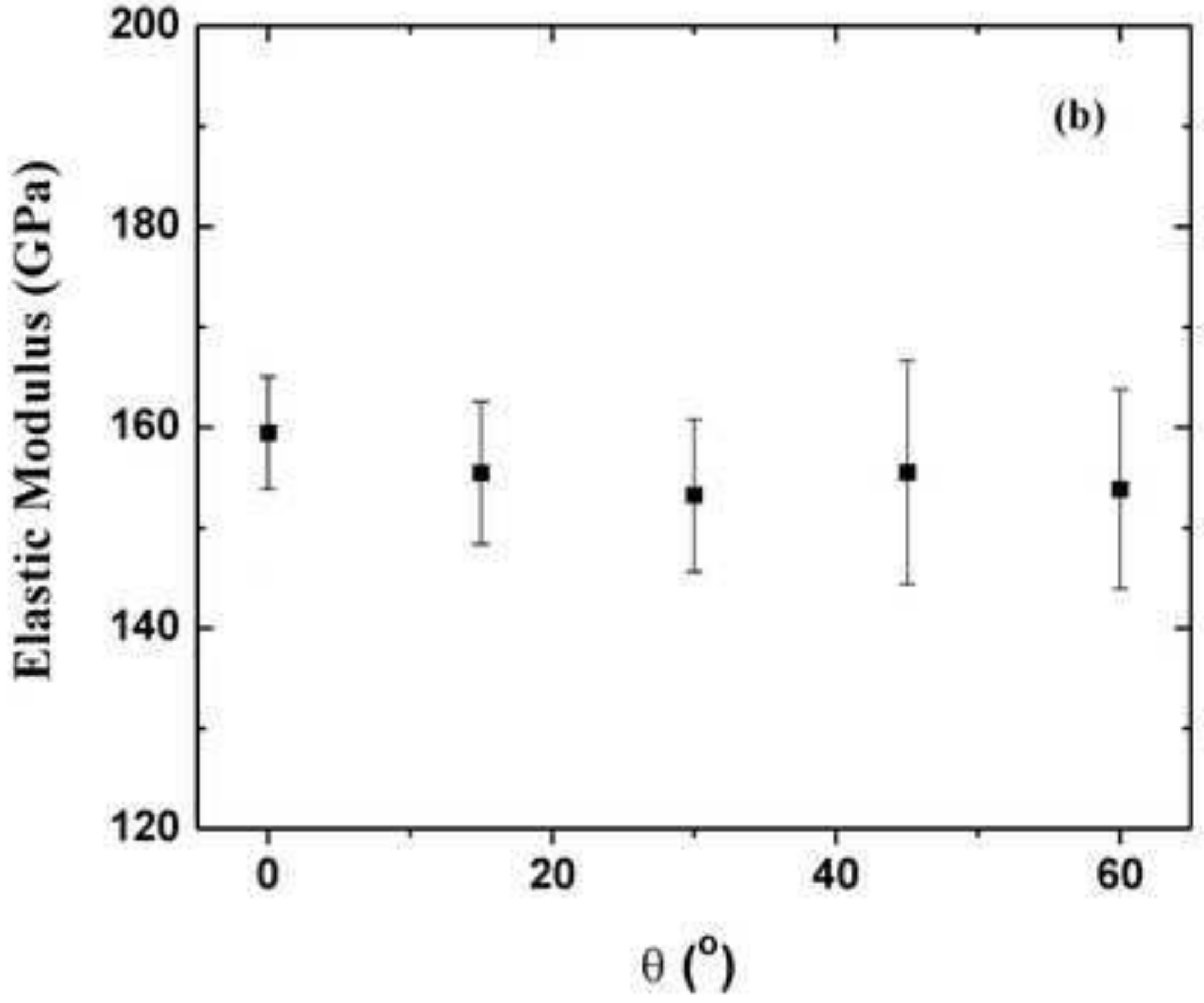

fig7

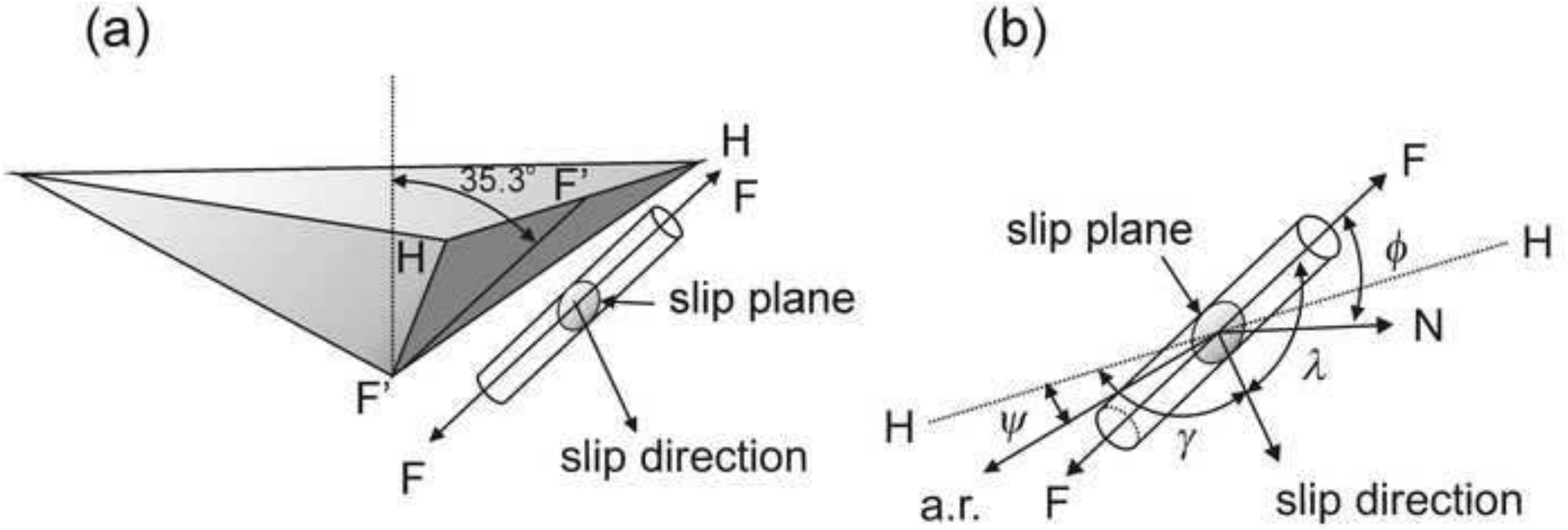

fig8
Click here to download high resolution image

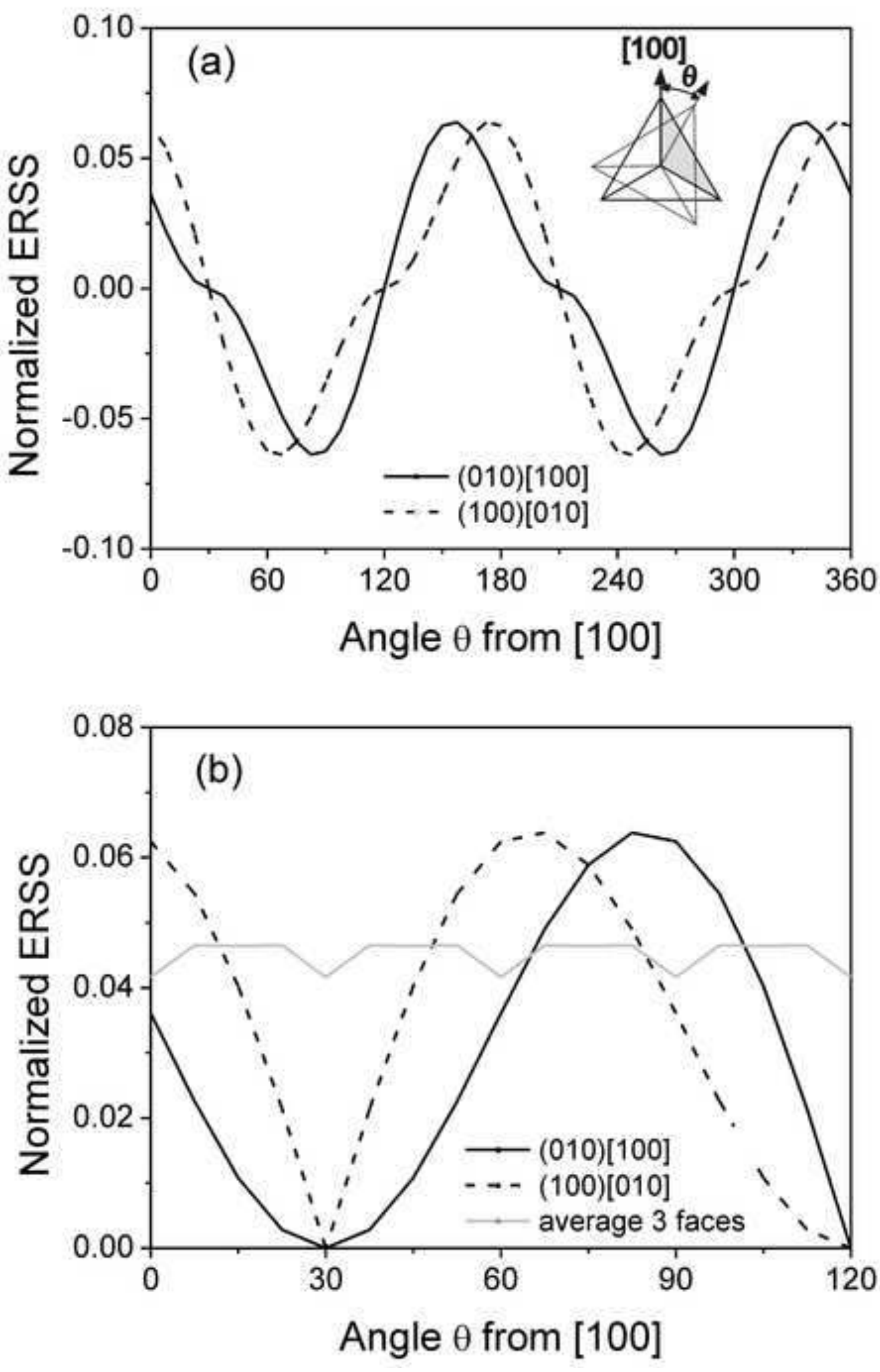